\documentclass[12pt]{iopart}
\usepackage{iopams}  
\usepackage{graphicx}
\usepackage{color}
\usepackage{hyperref}
\usepackage{tikz} 
\usepackage{url}
\usepackage{aas_macros}

\begin{document}

\title[Tidal deformability in the black hole compactness limit]{Where is Love? Tidal deformability  in the black hole compactness limit}

\author{Cecilia Chirenti$^{1,2,3,4}$, Camilo Posada$^5$ and
Victor Guedes$^6$}
\address{$^1$ Department of Astronomy, University of Maryland, College Park, \\MD 20742-2421, USA}
\address{$^2$ Astroparticle Physics Laboratory NASA/GSFC, Greenbelt, Maryland 20771, USA}
\address{$^3$ Center for Research and Exploration in Space Science and Technology, NASA/GSFC, Greenbelt, MD 20771, USA}
\address{$^4$ Center for Mathematics, Computation and Cognition, UFABC, Santo Andr\'e, SP  09210-170, Brazil}
\address{$^5$ Institute of Physics and Research Centre of Theoretical Physics and Astrophysics, Silesian University in Opava, CZ-746 01 Opava, Czech Republic}
\address{$^6$ Center for Natural and Human Sciences, UFABC, Santo Andr\'e, SP  09210-170, Brazil}
\ead{chirenti@umd.edu}

\begin{abstract}
One of the macroscopically measurable effects of gravity is the tidal deformability of astrophysical objects, which can be quantified by their tidal Love numbers. For planets and stars, these numbers measure the resistance of their material against the tidal forces, and the resulting contribution to their gravitational multipole moments. According to general relativity, nonrotating deformed black holes, instead, show no addition to their gravitational multipole moments, and all of their Love numbers are zero. In this paper we explore different configurations of nonrotating compact and ultracompact stars to bridge the compactness gap between black holes and neutron stars and calculate their Love number $k_2$.  We calculate $k_2$ for the first time for uniform density ultracompact stars with mass $M$ and radius $R$ beyond the Buchdahl limit (compactness $M/R > 4/9$), and we find that $k_2 \to 0^+$ as $M/R \to 1/2$, i.e., the Schwarzschild black hole limit. Our results provide insight on the zero tidal deformability limit and we use current constraints on the binary tidal deformability $\tilde{\Lambda}$ from GW170817 (and future upper limits from binary black hole mergers) to propose tests of alternative models.
\end{abstract}

%
%
%
%
%

\section{Introduction}

\label{sec:intro}

One of the classical manifestations of gravity at work is the phenomenon of tides. Interestingly, however, erroneous theories and expectations have been abundant in history. For instance, Galileo's theory of tides is famously wrong \cite{Buettner}.\footnote{It is also a case of bad scientific work eventually causing a big headache for its author \cite{Brecht}.} Exactly four centuries later, the world became aware of the first detection of gravitational waves from the merger of a binary black hole system, another (and much more spectacular) example of gravity at work \cite{2016PhRvL.116f1102A}. 

It took two more years for the detection of a gravitational wave event in which the tidal deformability of the astrophysical compact objects involved was constrained by the observations \cite{TheLIGOScientific:2017qsa}. The observed electromagnetic counterpart of this compact binary coalescence \cite{2017ApJ...848L..12A} allowed its confident identification as the merger of two neutron stars. This observation added to the existing constraints on the properties of matter inside neutron stars (see for example \cite{2020ApJ...888...12M} and also the more recent results from NICER  \cite{2019ApJ...887L..24M,2019ApJ...887L..21R}). A softer equation of state (EOS, which relates the pressure $p$ and the energy density $\epsilon$ of the stellar matter in the barotropic case) allows a given total mass $M$ of the star to be squeezed in a smaller total radius $R$ than a harder EOS, resulting in a smaller tidal deformation. Less compact stars can be more easily deformed.

Black holes have the maximum allowed compactness and have been shown to have zero tidal deformability in the nonrotating case and in some slowly rotating special cases (see, for example, \cite{Damour:2009vw,PhysRevD.80.084018,2012JHEP...02..010K,PhysRevD.91.104018,PhysRevD.92.024010,2015PhRvL.114o1102G}). This does not mean, however, that the black hole horizon is not deformed by tidal fields (it is). Instead, it results that an observer at infinity will not see a deviation from the asymptotic gravitational multipole moments of the black hole \cite{2015PhRvL.114o1102G}. As the black hole does not have an internal structure to be deformed, the distorted black hole still has no hair. Some aspects of this apparently counter-intuitive result have been considered  puzzling in an effective field theory approach to general relativity (GR) \cite{2016ForPh..64..723P}.

Therefore the expectation has been raised that \emph{only} black holes will have zero tidal deformability in GR, whereas all other alternative (and possibly exotic) structure models will have a likely very small but finite value. This observational signature has been conjectured as a possible test of such models using the inspiral part\footnote{Alternatively, a test using the \emph{ringdown} part of GW150914 was presented in \cite{2016PhRvD..94h4016C}.} of the waveform (see, for example,  \cite{PhysRevD.88.083002} for an early work and \cite{2017PhRvL.119i1101K} for a distinct theoretical prediction).

For different models of exotic compact objects (ECOs) it has been shown that their tidal Love numbers (see Section \ref{sec:general}) scale as $k\sim 1/\vert\log\xi\vert$, where $\xi=R/(2M) - 1$ is a parameter which indicates how close the object is to a black hole geometry \cite{2017PhRvD..95h4014C}. Compared to neutron stars, the tidal Love numbers for ECOs with Planckian corrections near the horizon are smaller by $4 - 5$ orders of magnitude. Such low values of $k$ are beyond the current LIGO sensitivity \cite{Abbott:2018exr}. 

Here we are interested in bridging the \emph{compactness gap} between neutron stars and black holes.  
We consider a recently proposed extension of the well-known analytical solution for uniform density stars (or Schwarzschild stars) \cite{Schwarzschild:1916ae}, allowing the extension of the solution to compactnesses higher than the Buchdahl limit $M/R = 4/9$ \cite{Buchdahl:1959B} (see Section \ref{SStar} for more details). This toy model provides a smooth transition from less compact stars to models approaching the black hole compactness \cite{Mazur:2015kia}. Moreover, it has been shown that, in the slow rotation approximation (following the prescription by Hartle \cite{1967ApJ...150.1005H}), its multipoles approach the Kerr values in the limit $R \to 2M$  \cite{Posada-Aguirre:2016qpz}. We show here that its tidal deformability also tends to zero in this limit.

This paper is organized as follows. In Section \ref{sec:tidal} we briefly review the formulation for calculating the tidal Love number $k_2$ for a uniform density star and present results obtained in a formulation of the equations that allows their calculation beyond the Buchdahl limit. In Section \ref{sec:obs} we present our results translated to the tidal deformability parameter $\bar{\Lambda}$ compared with current constraints for the neutron star EOS and discuss a test of alternative models using gravitational wave observations. Finally, we present our conclusions in Section \ref{sec:final}. Unless otherwise stated, we use $c = G = 1$ units.

\section{Tidal deformability}
\label{sec:tidal}
\subsection{General formulation}
\label{sec:general}
The tidal Love number quantifies the deformations of the quadrupole moments of a star induced, for example, by its companion in a binary system. The Love number $k_{2}$, or alternatively the tidal deformability $\Lambda$, provides a connection between the tidal fields $E_{ij}$ and the induced quadrupole moments $Q_{ij}$, given by \cite{Hinderer:2007mb,Damour:2009vw} 
\begin{equation}\label{quadrupole}
Q_{ij}=-\frac{2k_{2}R^5}{3}E_{ij}\equiv-\Lambda\,E_{ij}\,.
\end{equation}  
It is also conventional to introduce the dimensionless tidal deformability
\begin{equation}\label{tidal}
\bar{\Lambda}=\Lambda/M^5=2k_{2}/(3C^5)\,,
\end{equation} 
with $C\equiv M/R$ denoting the compactness of the configuration. Equation (\ref{tidal}) is most commonly employed in the context of the I-Love-Q relations for neutron stars \cite{Yagi:2013bca,Yagi:2013awa}.  Following the notation and formulation in \cite{Damour:2009vw}, we describe the unperturbed spacetime of a nonrotating star with the line element
\begin{equation}
\rmd s^2 = -\rme^{\nu(r)}\rmd t^2 + \rme^{\lambda(r)}\rmd r^2 + r^2\rmd\Omega^2\,,
\end{equation}
and the even-parity metric perturbation $H=H_{0}=H_{2}$ satisfies a single second-order differential equation given by
\begin{equation}\label{main}
\frac{\rmd^2H}{\rmd r^2}+C_{1}(r)\frac{\rmd H}{\rmd r}+C_{0}(r)H=0\,,
\end{equation}
where the coefficients $C_1$ and $C_0$ are 
\begin{eqnarray}
C_{1}(r) &=& \frac{2}{r}+\rme^{\lambda}\left[\frac{2m}{r^2}+4\pi r(p-\epsilon)\right]\,, \\
C_{0}(r) &=& \rme^{\lambda}\left[-\frac{l(l+1)}{r^2}+4\pi(\epsilon+p)\frac{\rmd\epsilon}{\rmd p}+4\pi(5\epsilon+9p)\right] 
- \left(\frac{\rmd\nu}{\rmd r}\right)^2\,.
\end{eqnarray}
This expression generalizes the result reported in \cite{Hinderer:2007mb} to any value of the multipole order $l$. Using the logarithmic derivative $h(r) \equiv (r/H)\rmd H/\rmd r$, eq. (\ref{main}) becomes the corresponding Riccati equation 
\begin{equation}
\label{eq:ricatti}
r\frac{\rmd h}{\rmd r}+h(h-1)+rC_{1}h+r^2C_{0}=0\,,
\end{equation}
with the regular solution near the origin
\begin{equation}
\label{eq:sol_series}
h(r) = l + \frac{4\pi\left[2l(l+1)(\epsilon_0/3)-(5\epsilon_{0}+9p_{0})\right]}{2l+3}r^2 + \mathcal{O}(r^4)\,,
\end{equation}
where $\epsilon_0$ and $p_{0}$ indicate the central values of the energy density and pressure, respectively. We can determine the Love number $k_{2}$ from the following expression
\begin{eqnarray}
k_{2}(C,h_{R})=\frac{8}{5}(1-2C)^2C^5[2C(h_R-1)-h_R+2]\nonumber\\
\times\{2C\big[4(h_R+1)C^4 + (6h_R-4)C^3 +(26-22h_R)C^2 \nonumber\\
+ 3(5h_R-8)C - 3h_R+6\big] +3(1-2C)^2\nonumber\\
\times\left[2C(h_R-1)-h_R+2\right]\log(1-2C)\}^{-1}.
\label{k2}
\end{eqnarray}
where $h_R$ is the surface value of $h$ at $r = R$.

\subsection{Uniform density stars}
\label{sec:uniform}

Uniform density stars are not a realistic model for neutron stars. The known pathologies of their description, like a density discontinuity at the surface and undefined sound speed are understood as the limiting case of an ultra-stiff EOS. Despite these shortcomings, they are useful toy models which can be described by an analytical solution of the Einstein field equations, as we briefly review below. 

In \cite{Hinderer:2007mb} the tidal Love number was computed for different neutron star models with realistic EOSs. The incompressible constant-density star, or Schwarzschild star, was considered by Damour and Nagar \cite{Damour:2009vw} and also more recently by \cite{Chan:2014tva}. 

To compute the tidal Love number for Schwarzschild stars, we follow the procedure described in \cite{Damour:2009vw}. We integrate eq. (\ref{eq:ricatti}) in the interior region of the star $0<r<R$ to find the value of the logarithmic derivative $h_R^{-}=h(R^{-})$. 

It is important to recall that in this case there is a $\delta$-function contribution which must be taken into consideration to obtain $h_R^{+}=y(R^{+})$ (as $h(r)$ is not continuous across $r = R$; see discussion in \cite{Damour:2009vw}). This correction is given by 
\begin{equation}\label{correction}
h_R^{+}=h_R^{-}-\left(\frac{4\pi R^3\epsilon}{M}\right)^{-}\,,
\end{equation}
which results in
\begin{equation}
h_R^+ = h_R^- - 3\,.
\end{equation}

The interior Schwarzschild solution describes a star of uniform total energy density $\epsilon$. It is usually given in terms of the auxiliary function $y$ (see, for example, \cite{Chandrasekhar:1974}) defined as
\begin{equation}
y^2=1-\left(\frac{r}{\alpha}\right)^2,\quad \textrm{with} \quad 
\alpha^2=\frac{3}{8\pi\epsilon} \equiv \frac{R^3}{R_{S}}\,,
\end{equation}
where $R$ is the radius of the star and $R_S \equiv 2M$ is the Schwarzschild radius. We note that $y$ is a decreasing function of $r$ that ranges from $y(0) = 1$ at the center to the surface value $y_1 \equiv y(R)$, which depends on the compactness of the star ($y_1$ is smaller for more compact stars).

The metric functions are given analytically inside the star by the expressions
\begin{equation}
\rme^\nu=\frac{1}{4}(3y_{1}-y)^2\quad \textrm{and} \quad \rme^{\lambda}=\frac{1}{y^2}\,,\\
\end{equation}
which match continuously to the (exterior) Schwarzschild metric outside the star. The fluid pressure $p$ inside the star is also given analytically by
\begin{equation}
\label{eq:p(y)}
p=\epsilon\left(\frac{y-y_{1}}{3y_{1}-y}\right)\,,
\end{equation}
and $p \to 0$ as $r \to R$. For the central pressure to be positive and finite everywhere inside the star, the Buchdahl condition  $R/R_{\rm S}>9/8$ (or $3y_{1}>1$ in this parametrization) must be satisfied \cite{Buchdahl:1959B}.

\subsection{Ultracompact Schwarzschild stars}
\label{SStar}
The well-known interior Schwarzschild solution discussed in the previous Section \ref{sec:uniform} describes uniform density stars of increasing compactness up to the Buchdahl limit $M/R = 4/9$, or $R = (9/8)R_S$. An extension of this solution beyond this limit was recently found in \cite{Mazur:2015kia}. It allows us to describe uniform density stars in the range $R_{S}<R<(9/8)R_{S}$, that is, approaching the Schwarzschild black hole compactness limit, where it essentially becomes a gravastar \cite{Mazur:2001fv,Mazur:2004fk}.

Properties of the ultracompact Schwarzschild stars have been explored in some follow-up works. It was found that, in the slowly rotating approximation, their multipole moments approach the Kerr values for increasing compactness \cite{Posada-Aguirre:2016qpz}, and that these solutions are radially stable \cite{Camilo:2018goy}, in the sense that the critical adiabatic index of the perturbed fluid required for stable radial oscillations becomes finite beyond the Buchdahl limit, extending the classical radial stability analysis of \cite{1964ApJ...140..417C}. It is interesting to note that the analysis of axial perturbations did not show echoes in the waveform \cite{2019PhRvD.100d4027K}, as proposed for generic exotic compact objects \cite{2016PhRvL.116q1101C}. A time-dependent model that allows for the evolution from a low compactness star to a gravastar in the black hole compactness limit was presented in \cite{2019PhRvD..99h4021B}.

In order to extend the solution of eq. (\ref{eq:ricatti}) beyond the Buchdahl limit, it is convenient to introduce a new coordinate $x$ as 
\begin{equation}\label{xcoord}
x \equiv 1-y = 1-\sqrt{1-\left(\frac{r}{\alpha}\right)^2}\,,
\end{equation}
which ranges from $x(0) = 0$ to the surface value $x_1 = 1-y_1\equiv x(R)$, also depending on the compactness of the star ($x_1$ is larger for more compact stars) \cite{Chandrasekhar:1974}. The pressure (\ref{eq:p(y)}) can now be written as
\begin{equation}
p=\epsilon\left(\frac{1-x-y_1}{k+x}\right)
\label{eq:p(x)}
\end{equation}
where $k$ is a constant defined as
\begin{equation}\label{k}
k\equiv3y_{1}-1\,.
\end{equation}
As we can see from (\ref{eq:p(x)}), the pressure diverges at $x_0 \equiv -k$ inside the star. Mazur and Mottola \cite{Mazur:2015kia} have applied a generalization of the Israel junction conditions \cite{Israel:1966rt} to the null boundary layer at $x_0$, obtaining an anisotropic pressure contribution to the stress energy tensor at the hypersurface, given by 
\begin{equation}
8\pi \rme^{\frac{\nu+\lambda}{2}}\alpha^2[1-(1-x)^2]^2(p_t - p) = \frac{8\pi\epsilon}{3}R^3_0\delta(x - x_0)\,,
\label{eq:pt}
\end{equation} 
where $p_t$ is the tangential pressure. This term results in a physical surface tension associated with a positive integrable transverse pressure contribution to the Komar mass, see \cite{Mazur:2015kia,Posada-Aguirre:2016qpz,2019PhRvD..99h4021B}. This is the key result that allows the ultracompact interior Schwarzschild solution to be interpreted in physical terms. The pressure anisotropy should be taken into account in the calculation of the tidal deformability. However, as we can see from eq. (9) of \cite{2019PhRvD..99j4002B}, the correction to the metric perturbation eq. (\ref{main}) comes in the form of an additional term proportional to $\rmd p_t/\rmd p$. For eq. (\ref{eq:pt}) we have $\rmd p_t/\rmd p = 1$ and the isotropic form of the perturbation equation (\ref{main}) considered here is recovered.

In terms of $x$ (\ref{xcoord}), $h(x) = [x(2-x)/(1-x)](1/H)\rmd H/\rmd x$ and eq. (\ref{eq:ricatti}) becomes
\begin{equation}\label{ricattix}
\frac{\rmd h}{\rmd x} + \frac{(1-x)}{x(2-x)}h(h-1)+\frac{\alpha(1-x)}{\sqrt{x(2-x)}}C_{1}(x)h 
+ \alpha^2(1-x)C_{0}(x)=0,
\end{equation}
with the coefficients $C_1$ and $C_0$ written explicitly in terms of $x$ as
\begin{equation}\label{C1x}
C_{1}(x) = \frac{2}{\alpha\sqrt{x(2-x)}}+\frac{\sqrt{x(2-x)}}{\alpha(1-x)^2}\left(\frac{1-k-2x}{k+x}\right), 
\end{equation}
\begin{equation}
C_{0}(x) = \frac{1}{\alpha^2(1-x)^2}\left[\frac{-l(l+1)}{x(2-x)}+\frac{3(k-2x+3)}{k+x}\right] 
- \frac{4x(2-x)}{[\alpha\,(k+x)\,(1-x)]^2}.
\label{C0x}
\end{equation}

\subsection{Implementation and numerical results}
\label{sec:results}

Eq. (\ref{ricattix}) has a regular series solution near the origin of the form

\begin{equation}\label{hxc}
h(x) = l + \left(2a_1 + \frac{l}{2}\right)x + \mathcal{O}(x^2).
\end{equation}
where $a_1$ is given by a standard application of the Frobenius method, and we have
\begin{equation}
a_1 = \frac{3k-22}{14k}\quad \textrm{for} \quad {l = 2}.
\end{equation}
Therefore expression (\ref{hxc}) implies a finite nonzero value for $h'(x)$ at $x = 0$ (compare with eq.(\ref{eq:sol_series})), which is, however, just an artifact of our coordinate transformation. The physical condition to be satisfied is regularity of $H(r)$ at $r = 0$ (or $\rmd H/\rmd r = 0$ at $r = 0$), which does not require $\rmd h/\rmd x = 0$ at $x = 0$. 

\begin{figure}[htb!]
	\centering
	\includegraphics[scale = 0.7]{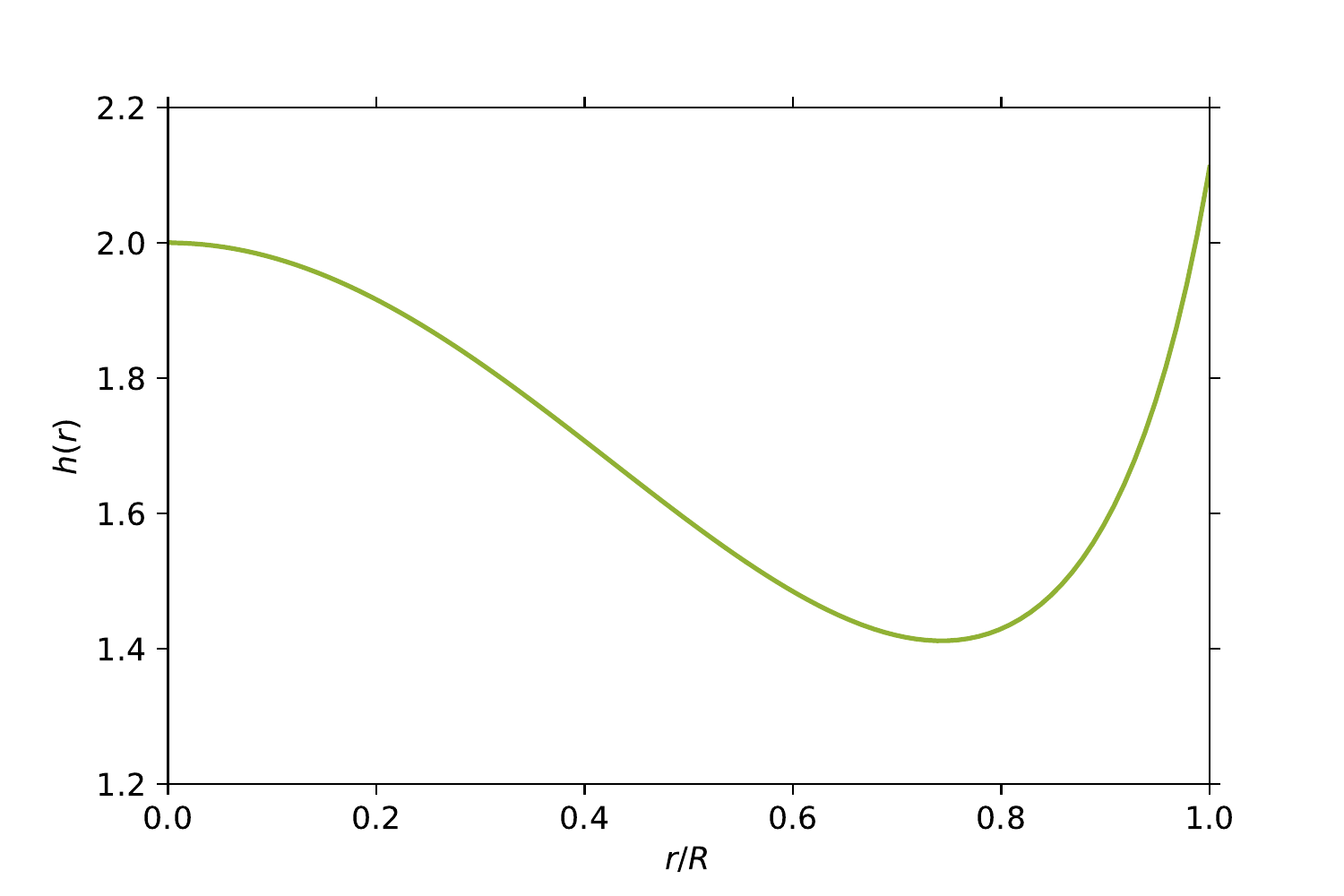}
	\includegraphics[scale = 0.7]{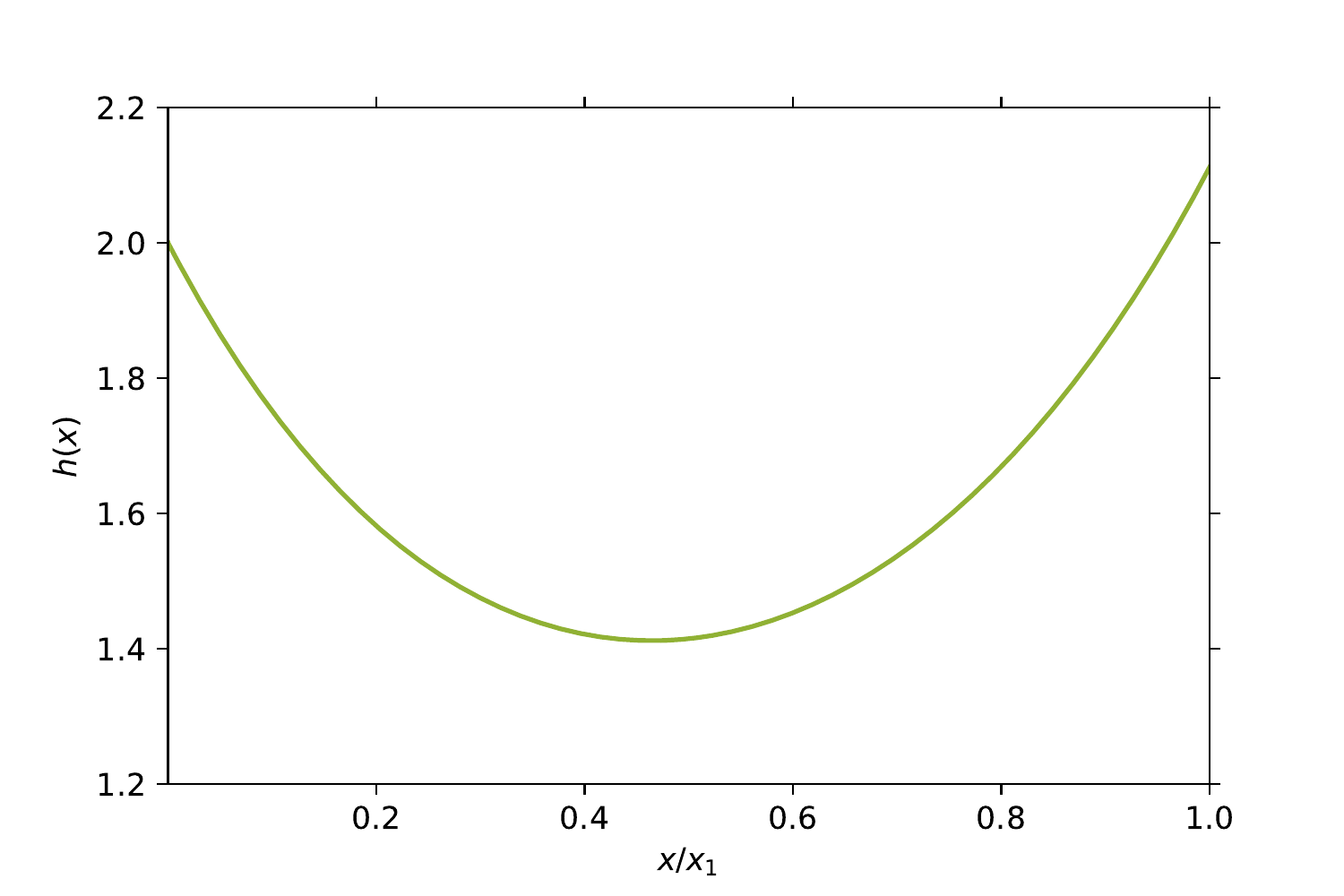}
	\caption{Example of the solutions $h(r)$ (eq.(\ref{eq:ricatti})) in the upper panel and $h(x)$ (eq.(\ref{ricattix})) in the lower panel, for $l = 2$ and a uniform density star with $R/R_S = 1.3$. The solutions look very different as a result of the coordinate transformation, but $h(x(\bar{r})) = h(\bar{r})$ at every point $r = \bar{r}$ inside the star. The formulation in terms of the $x$ coordinate allows us to go beyond the Buchdahl limit and consider stars with compactness approaching $M/R = 1/2$.}
	\label{fig:sols}
\end{figure}

An example of this effect is shown in Figure \ref{fig:sols}, where we present the same solution with $l = 2$ obtained in $r$ by solving eq. (\ref{eq:ricatti}) (top panel) and in $x$ by solving eq. (\ref{ricattix}) (bottom panel). For this comparison we choose $R/R_S = 1.3$ (less compact than the Buchdahl limit) where both formulations are valid. Although the curves $h(r)$ and $h(x)$ look rather different, they produce the same value $h^-_R$ (needed for the calculation of the Love number $k_2$), that is, $h(R) = h(x_1)$.

As a further application of our formulation and a test of our code, we are able to determine the value of $k_2$ at the Buchdahl limit. We find $k_2^{\rm Buchdahl} = 0.0017103$, which agrees with the analytically estimated result reported by \cite{Damour:2009vw}.

\begin{figure}[htb!]
	\centering
	\includegraphics[scale = 0.5]{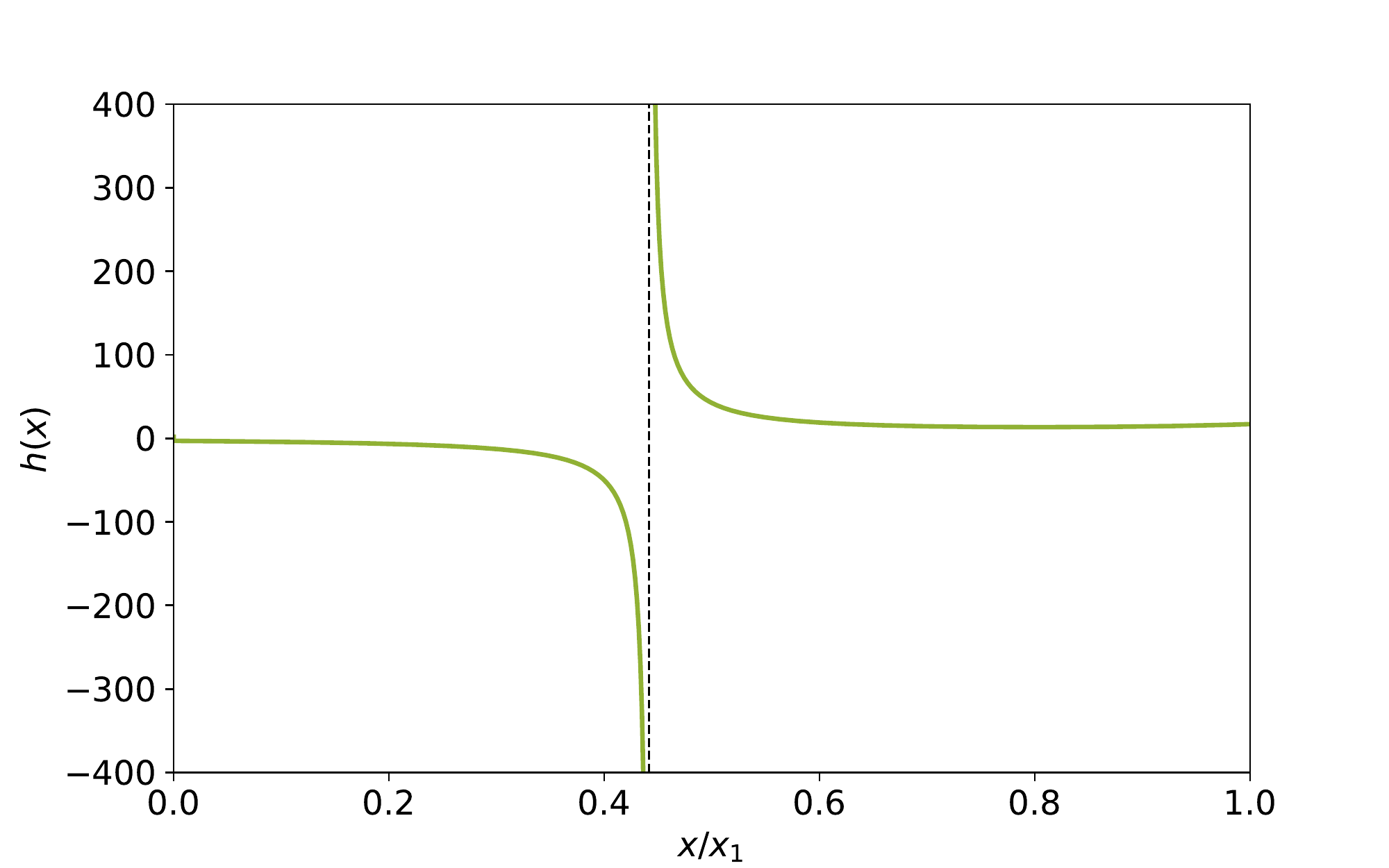}
	\includegraphics[scale = 0.5]{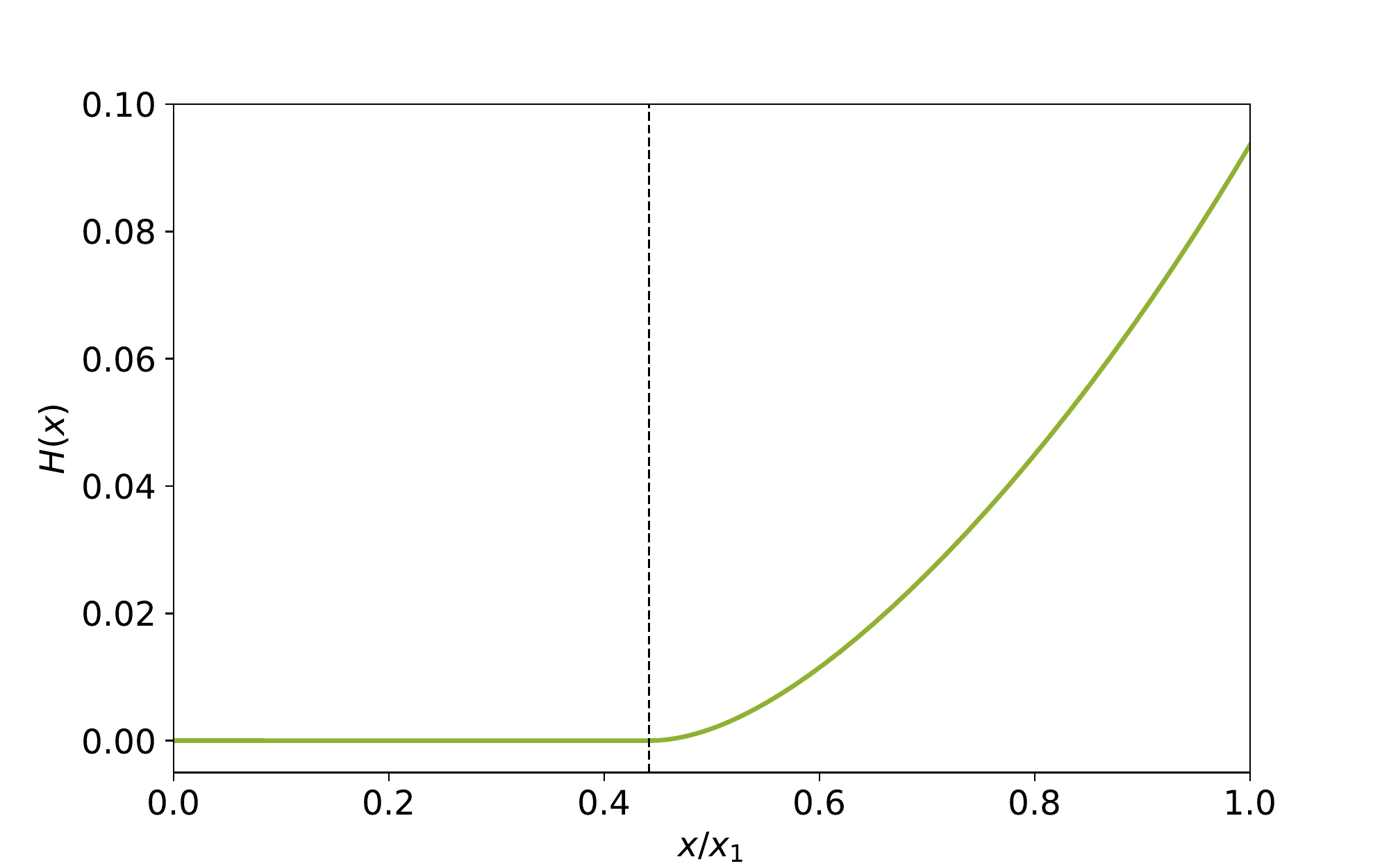}
	\caption{Example of the solutions $h(x)$ (upper panel) and the corresponding $H(x)$ (lower panel), for $l = 2$ and a ultracompact uniform density star with $R/R_S = 1.05$. The vertical dotted line indicates the singular point $x_0 = -k$ for $h(x)$, which is a regular singular point for $H(x)$ (see the series expansions (\ref{eq:series_h(-k)}) and (\ref{eq:series_H(-k)}), respectively). The results for the tidal deformability obtained with both numerical implementations in the case shown here agree within 0.01\%.}
	\label{fig:sols_beyond}
\end{figure}

Finally, we extend our results to the compactness gap between the Buchdahl limit and the Schwarzschild black hole. In this compactness range, eq. (\ref{ricattix}) has another singular point at $x_0 = -k > 0$. Eq. (\ref{ricattix}) has a singular solution at $x_0 = -k$ given in terms of the Laurent series
\begin{equation}
\label{eq:series_h(-k)}
h(x) = -\frac{2k(k+2)}{k+1}\left(\frac{1}{x+k}\right) + \frac{1-k(k+2)}{(k+1)^2} + \mathcal{O}(x+k)\,,
\end{equation}
and an example of the solution for $h(x)$ beyond the Buchdahl limit can be seen in the upper panel of Figure \ref{fig:sols_beyond}. The corresponding solution for $H(x)$, obtained from eq. (\ref{main}) using the coordinate transformation (\ref{xcoord}), is shown for comparison and validation of our results in the lower panel, where we can see the regular behavior of $H(x)$ at $x_0 = -k$. In fact, we can show that
\begin{equation}
\label{eq:series_H(-k)}
H(x) = b_0(x+k)^2\left[1 + \frac{3(k+1)}{k(k+2)}(x+k) + \mathcal{O}(x+k)^2 \right]\,,
\end{equation}
where $b_0$ is a constant fixed by the numerical integration of $H(x)$ in the interval $0<x<x_0$. The results obtained from the numerical solutions for $h(x)$ and $H(x)$ agree within less than 0.1\% for $M/R < 0.491$. For higher values of the compactness, the agreement is still very good, but numerical errors make it harder to estimate the true relative error.

\begin{figure}[htb!]
	\centering
	\includegraphics[width = 0.7\linewidth]{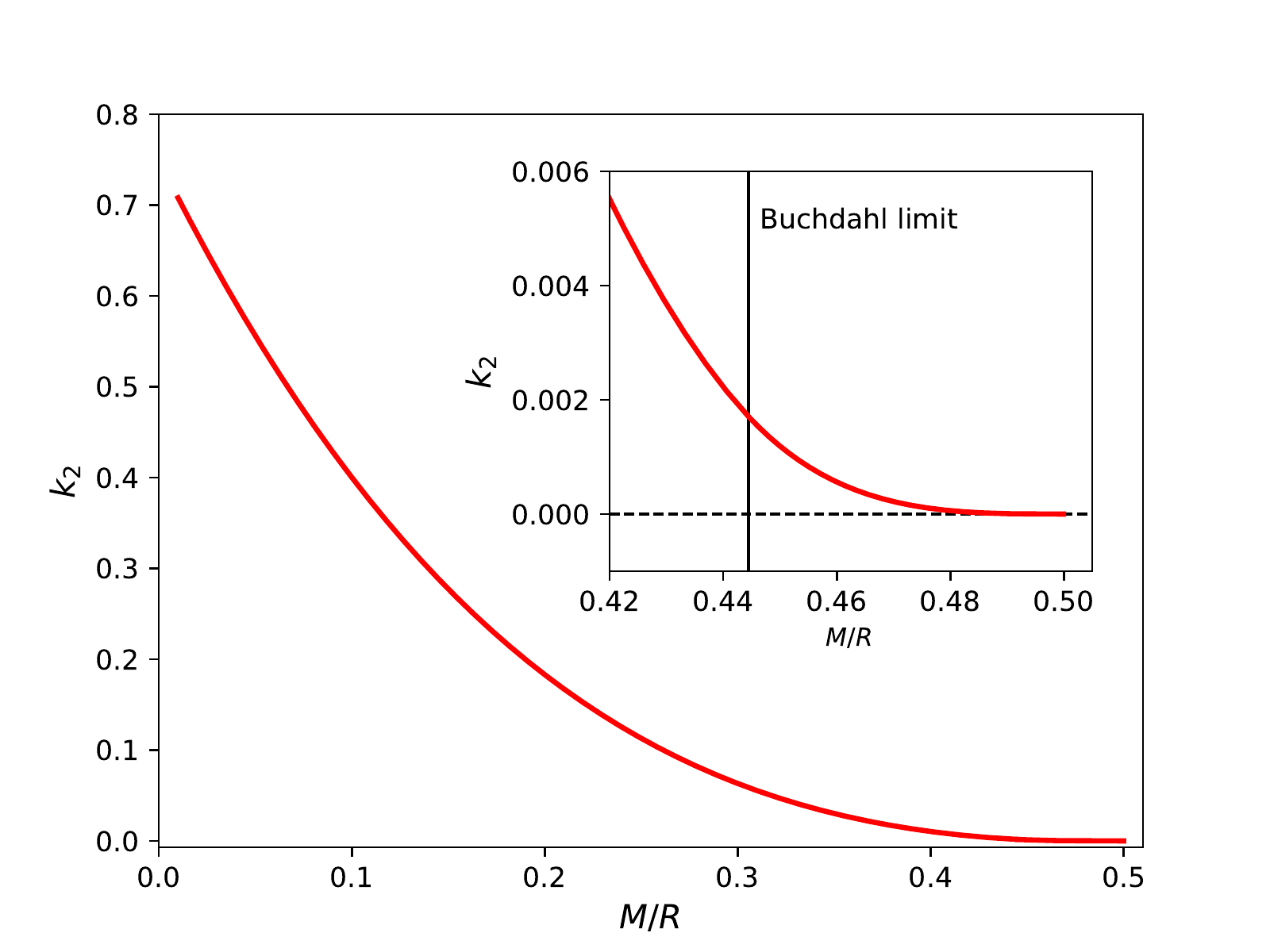}
	\caption{Tidal Love number $k_2$ as a function of the compactness for uniform density stars below and above the Buchdahl limit $R/R_S = 1.125$. Note that $k_2$ is continuous and approaches zero smoothly as $M/R\to 1/2$.}
	\label{fig:k2}
\end{figure}

As we can see from Figure \ref{fig:k2}, $k_2$ is a continuous function of $M/R$ across this limit. The inset shows the smooth continuation of $k_2$ as the Buchdahl limit is crossed. Stars more compact than this limit show a central region of negative pressure from $x = 0$ to $x_0 = -k$, with $x_0 = 0$ for the Buchdahl limit and $x_0 \to x_1$ as $M/R \to 1/2$ \cite{Mazur:2015kia}. 

\begin{figure}[htb!]
	\centering
	\includegraphics[width = 0.8\linewidth]{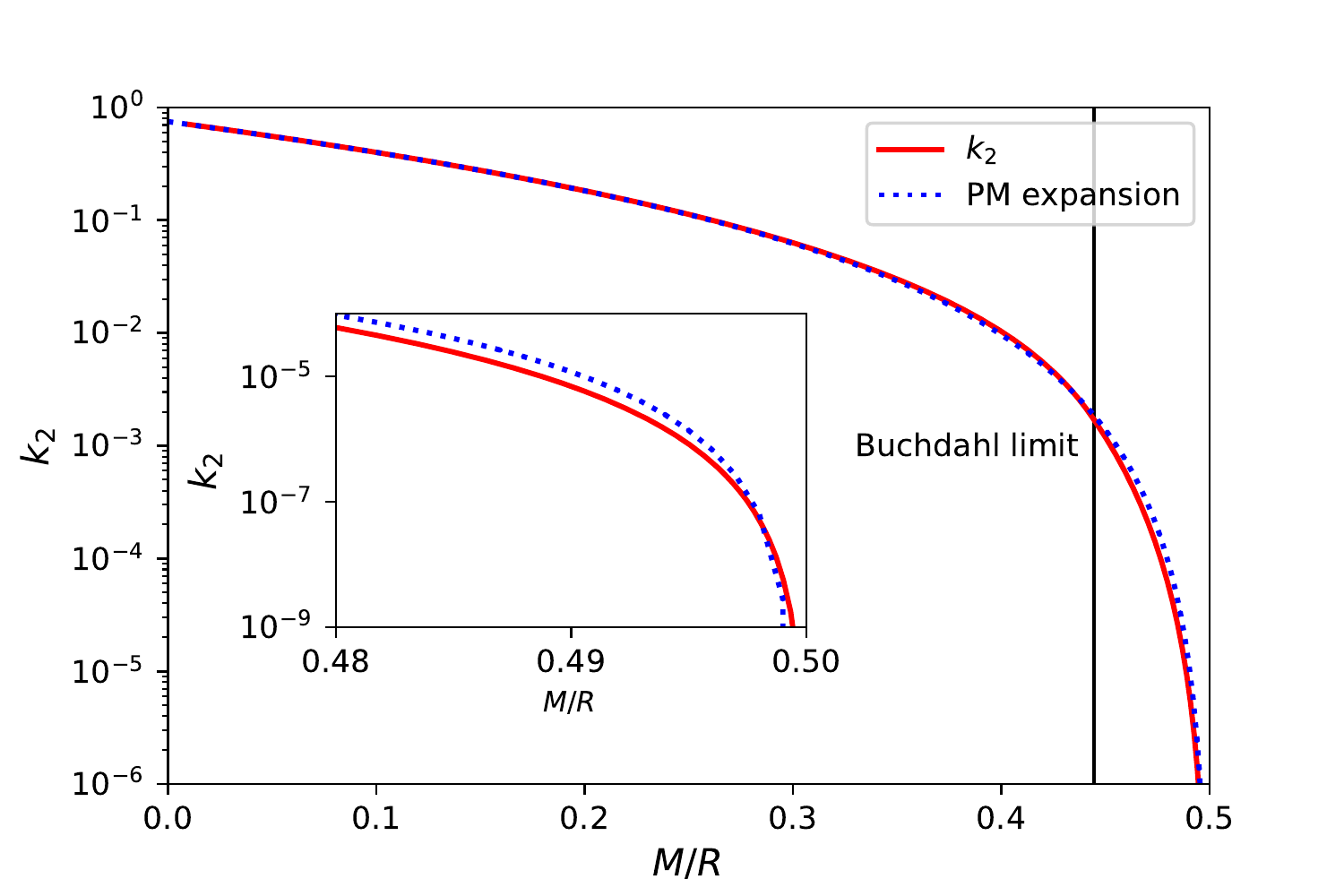}
	\caption{Same as Figure \ref{fig:k2}, but in logscale to show the approach to the black hole compactness limit. We also compare our numerical results with the Post-Minkowskian (PM) approximation given in \cite{Chan:2014tva} to 6th order in the compactness.}
	\label{fig:logk2}
\end{figure}

In Figure \ref{fig:logk2} we show in the log-scale how $k_2$ approaches 0 as $M/R$ approaches $1/2$, and compare our results with the Post-Minkowskian (PM) approximation presented in \cite{Chan:2014tva}. We can see that the PM approximation (given to 6th order in $M/R$) agrees very well with our numerical results up to $M/R \sim 0.4$, but the agreement becomes worse beyond the Buchdahl limit, as expected.

It is important to note that \cite{Pani:2015tga} found a \emph{negative} tidal deformability for the infinitesimally thin shell gravastar, also approaching zero at the black hole limit. Although the ultracompact Schwarzschild star essentially becomes a gravastar in the same limit, lower compactness solutions from both models are very different. Beyond the Buchdahl limit, the Schwarzschild star begins to show an increasing core of de Sitter space, surrounded by matter with positive pressure. The thin shell gravastar, however, is entirely filled by the negative pressure fluid leading to this unusual behavior in the tidal deformability (see discussion in \cite{Pani:2015tga}). \footnote{Anisotropic stars and some other exotic models have also been shown to have negative tidal deformability \cite{2016CQGra..33i5005Y,2017PhRvD..95h4014C,2019PhRvD..99j4072R}.}

From these results we can conclude that $k_2 \to 0$ does not require a black hole or the existence of an event horizon (although $k_2 = 0$ is only attained in the black hole configuration). Rather, this limit seems to be a consequence of $M/R \to 1/2$. This result (for non-black holes) may be at least partially understood in intuitive physical terms if we remember that the acceleration $a(r) = -\rme^{-\lambda/2}\lambda'/2 $ of a stationary particle held at a distance $r$ from a black holes diverges as $r \to 2M$. Therefore we can expect that in our case, as the compactness approaches $M/R = 1/2$, it will eventually become impossible to deform the shape of the star beyond its already tidally deformed equipotential surface.

\section{Observational implications}
\label{sec:obs}

In this Section we translate our results to $\bar{\Lambda}$ in order to include current constraints on the tidal deformability obtained from the first detection of a binary neutron star merger by LIGO and Virgo, GW170817 \cite{2018PhRvL.121i1102D,2019PhRvX...9a1001A}. 

A binary neutron star will merge sooner than a binary black hole with comparable masses due to the   energy required to tidally deform the stars. The leading order contribution to the inspiral waveform is proportional to the binary tidal deformability $\tilde{\Lambda}$ \cite{PhysRevD.77.021502}. This is a mass-weighted average of the dimensionless tidal deformabilities $\bar{\Lambda}_1$ and $\bar{\Lambda}_2$ of the individual stars in the binary with total masses $M_1$ and $M_2$, respectively. It is given by
\begin{equation}
\tilde{\Lambda} = \frac{16}{13}\frac{(M_1 + 12M_2)M_1^4\bar{\Lambda}_1 + (M_2 + 12M_1)M_2^4\bar{\Lambda}_2}{(M_1+M_2)^5}\,,
\end{equation}
which results in $\tilde{\Lambda} = \bar{\Lambda}_1 = \bar{\Lambda}_2$ for an equal mass binary, that is, if $M_1 = M_2$. The inferred constraint from GW170817 is $50 \lesssim \tilde{\Lambda} \lesssim 800$ at the 90\% confidence level \cite{2019PhRvX...9a1001A}. As the mass ratio $q = m_2/m_1$ for the event is compatible with equal masses, we will take $\tilde{\Lambda} = \bar{\Lambda}$ in our analysis.

\begin{figure}[htb!]
	\centering
	\includegraphics[width = 0.7\linewidth]{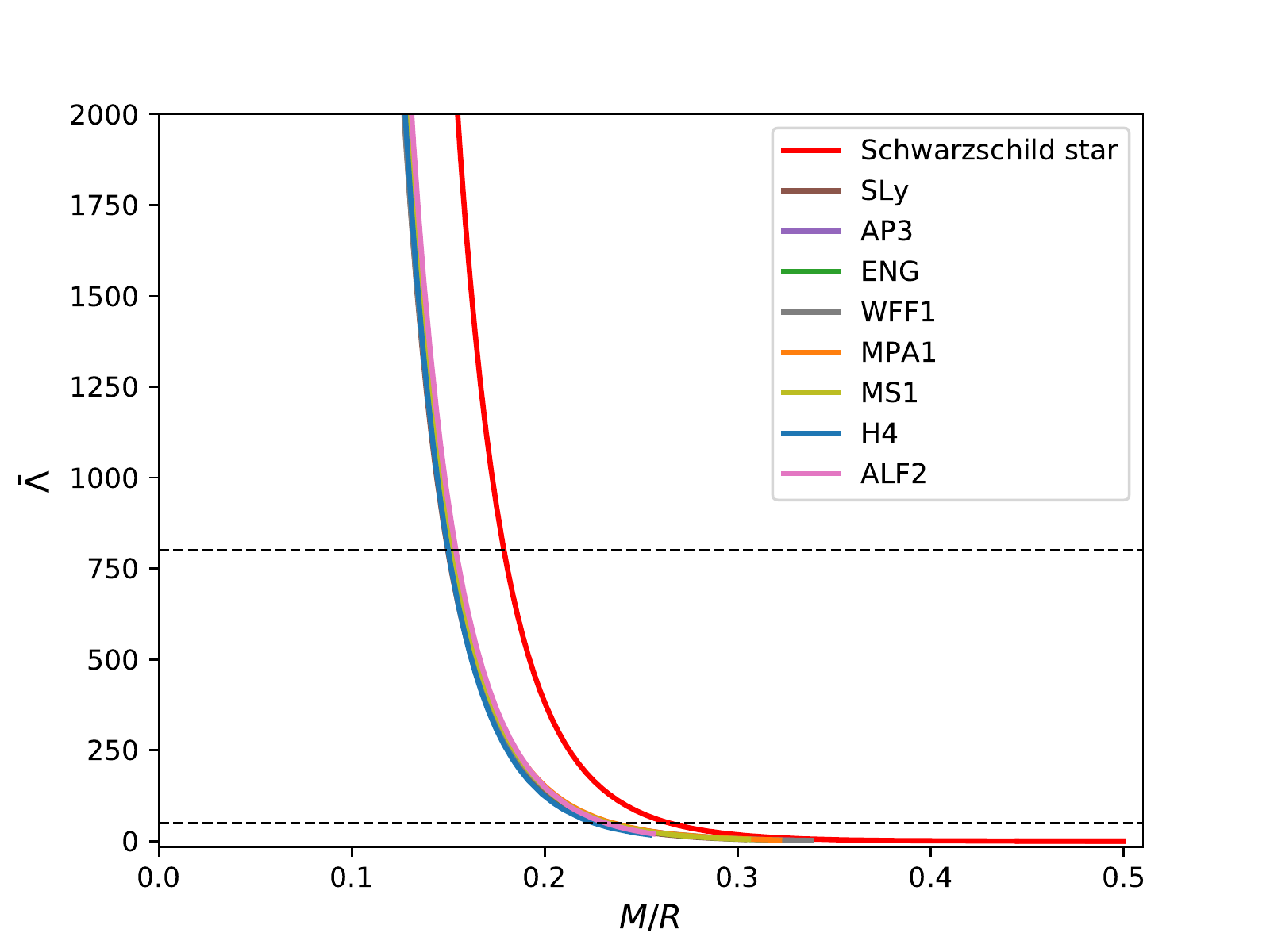}
	\includegraphics[width = 0.7\linewidth]{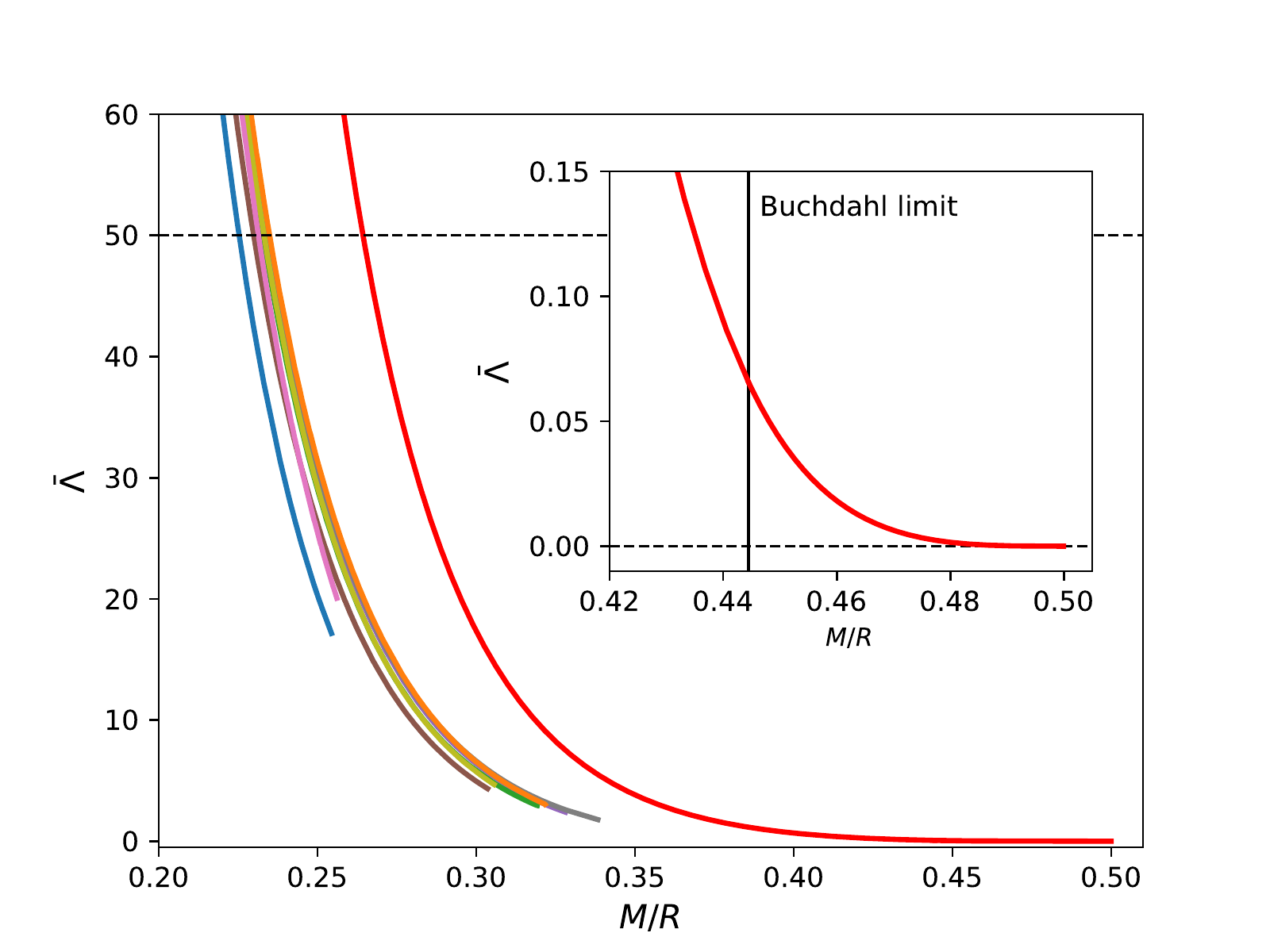}
	\caption{Upper panel: Tidal deformability $\bar{\Lambda}$ for the uniform density Schwarzschild star and neutron stars with 8 different realistic EOSs using the piecewise polytropic parametrization proposed by \cite{2009PhRvD..79l4032R}. The upper and lower LIGO/Virgo constraints from GW170817 are shown by the dotted lines at $\bar{\Lambda} = 800$ and 50, respectively. Lower panel: Same as the upper panel, but zooming in on higher compactnesses. The black vertical line shows the Buchdahl limit.}
	\label{fig:sols_ns}
\end{figure}

In Figure \ref{fig:sols_ns} we compare our results with 8 different realistic EOSs for neutron stars. The Schwarzschild star always has larger $\bar{\Lambda}$ than a neutron star of the same compactness, due to its mass distribution. This difference decreases with increasing compactness, and it is almost a factor 2 for the most compact neutron stars, at $M/R \sim 0.34$. 

Conversely, the same tidal deformability, as inferred by gravitational wave detections, would be compatible with a neutron star of some compactness or with a more compact Schwarzschild star. Assuming that their masses are the same, the gravitational wave signature would be distinguishable only by the maximum inspiral frequency before merger, which would be higher for the more compact object. 

Taking as an example ${\tilde\Lambda}=50$, which is the 5th percentile reported by \cite{2019PhRvX...9a1001A}, the corresponding neutron star would have $M/R \sim 0.23$, whereas the Schwarzschild star would have $M/R \sim 0.26$. This small difference would lead to a maximum gravitational wave frequency during the inspiral $f_{\rm max} \sim \sqrt{GM/(\pi^2R^3)}$ that would be $20\%$ larger in the case of the Schwarzschild star. For an equal mass binary with masses $M_1 = M_2 = 1.4 M_{\odot}$, it would mean a change from 1.2 kHz to 1.5 kHz. (Unfortunately LIGO and 
Virgo have poor sensitivity at either frequency.)

More interestingly, the Schwarzschild star's tidal deformability goes to lower values than any realistic neutron star EOS once its compactness is higher than approximately 0.38, approaching zero as $M/R \to 0.5$, as we saw in Section \ref{sec:results} for the Love number $k_2$. Currently, all data analyses of binary black hole mergers have assumed zero tidal deformability. However, the inclusion of the tidal deformability as an extra parameter would provide an upper limit for $\bar{\Lambda}$, which could be used to constrain the compactness of exotic black hole alternative models (see \cite{2018arXiv180408026J} for a constraint on boson star parameters using simulated data). 

Let us assume an equal mass black hole binary with $M_1 = M_2 = 30 M_{\odot}$ and a hypothetical observational constraint $\bar{\Lambda} < 50$. The frequency $f_{\rm max}$ of the uniform density model with $\bar{\Lambda} = 50$ would be approximately 100 Hz, nearly 40\% lower than the approximtely 270 Hz expected for the black hole case. These maximum chirp frequencies should be discernible in the LIGO band (with higher sensitivity in the range $50-500$ Hz) for a sufficiently high signal-to-noise ratio (SNR).

\section{Conclusions}
\label{sec:final}

The ongoing detections of gravitational waves allow us to start constraining possible deviations from general relativity and the existence of alternative models. So far, there are no clear indications for the need of corrections to Einstein's theory. One of its predictions is that black holes have zero tidal deformability, and the analysis of gravitational wave events consistent with binary black hole mergers has used waveforms without the 5PN tidal contribution.

We calculated here for the first time the tidal deformability of a sequence of nonrotating uniform density stars crossing the Buchdahl limit and bridging the compactness gap to the black hole limit. The  Love number of these stars is a continuous function of the compactness, positive-valued and vanishing as $M/R \to1/2$. We propose that $k_2 \to 0$ is a consequence of the approach to the black hole compactness, but independent of the actual existence of a black hole or an event horizon.  

An analysis of available gravitational wave data allowing for nonzero tidal deformability in the binary black hole merger events would likely provide interesting upper limits for the black hole tidal deformability. These upper limits could be used to constrain the compactness of alternative models. Additionally, events with suitable initial masses, such as GW150914, will have a maximum inspiral frequency well inside the LIGO band, and the analysis of high SNR events could lead to promising constraints.

It has recently been suggested that the tidal deformability of a Kerr black hole may not vanish in general \cite{2020arXiv200700214L}. Although LIGO's current constraints on the binary black hole spins are still reasonably broad and mostly compatible with zero, an interesting formal extension of our work would be to follow the approach of \cite{2015PhRvD..92l4003P} and test whether the tidal deformability of ultracompact rotating uniform density stars also approach the nonzero value for rotating black holes derived by \cite{2020arXiv200700214L}.

\bigskip
\ack
We would like to thank Cole Miller and Eric Poisson for useful discussions. We are also grateful to Emil Mottola and John Miller for comments on an earlier version of this manuscript. CP acknowledges the support of the Institute of Physics and its Research Centre of Theoretical Physics and Astrophysics at the Silesian University in Opava. VG acknowledges support from the Brazilian National Council for Scientific and Technological Development (CNPq) and the S\~ao Paulo Research Foundation (FAPESP) through grant 2019/20740-5. CC acknowledges support from CNPq grant 303750/2017-0, the Simons Foundation through the Simons Foundation Emmy Noether Fellows Program at Perimeter Institute and NASA under award number 80GSFC17M0002. Research at Perimeter Institute is supported in part by the government of Canada through the Department of Innovation, Science and Economic Development and by the Province of Ontario through the Ministry of Colleges and Universities.

\section*{References}
\bibliographystyle{unsrt.bst}
\bibliography{references}

\end{document}